\documentclass[prl,twocolumn,showpacs,groupedaddress,citeautoscript]{revtex4}
\usepackage{graphicx}
\usepackage{amsmath}

\renewcommand{\L}{\mathrm{L}}
\newcommand{\R}{\mathrm{R}}
\newcommand{\eff}{\mathrm{eff}}
\renewcommand{\section}[1]{{\par\it #1.---}}

\newcommand{\void}[1]{}
\begin{document}
\title{
Non-adiabatic electron pumping: maximal current with minimal noise
}
\author{Michael Strass}
\author{Peter H\"anggi}
\author{Sigmund Kohler}
\affiliation{Institut f\"ur Physik, Universit\"at Augsburg,
        Universit\"atsstra{\ss}e~1, D-86135 Augsburg, Germany}
\date{22 September 2005}
\begin{abstract}

The noise properties of pump currents through an open double quantum
dot setup with non-adiabatic ac driving are
investigated.  Driving frequencies close to the internal
resonances of the double dot-system mark the optimal working points at
which the pump current assumes a maximum while its noise power possesses
a remarkably low minimum.  A rotating-wave approximation provides
analytical expressions for the current and its noise power
and allows to optimize  the noise characteristics.
The analytical results are compared to numerical
results from a Floquet transport theory.

\end{abstract}
\maketitle

In mesoscopic conductors, a cyclic adiabatic change of the
parameters can induce a pump current, i.e., a non-vanishing
dc current flowing even in the absence of any external bias voltage
\cite{Kouwenhoven1991a, Pothier1992a, Switkes1999a}.
For adiabatic quantum pumps \cite{Brouwer1998a, Altshuler1999a,
Wagner1999a, Moskalets2004a, Brandes2002a}, the transfered charge per
cycle is determined by the area enclosed in parameter space during the
cyclic evolution \cite{Brouwer1998a, Altshuler1999a}.  This implies
that the resulting current is proportional to the driving frequency
and, thus, suggests that \textit{non-adiabatic} electron pumping
is more effective.
For practical applications, it is also desirable to operate the
quantum pump in a low-noise regime.
It has been found that adiabatic pumps can be practically noiseless
\cite{Avron2001a}.  This happens, however, on the expense of
acquiring a small or even vanishing current \cite{Polianski2002a}.
Therefore, the question arises whether it is possible to boost the pump
current by increasing the driving frequency while keeping the noise
level very low.

Non-adiabatic electron pumping can be achieved experimentally with
double quantum dots under the influence of microwave radiation
\cite{VanderWiel1999a, VanderWiel2003a, Platero2004a, Cota2005a}.
In this letter we study the transport properties in this non-adiabatic
regime.  Our main aim is to find ideal parameter regimes in
which a large pump current is associated with low current noise.
For the optimization of the system parameters, it is beneficial to
obtain, besides a numerical solution, also analytic expressions for
the transport quantifiers.  Therefore, within a rotating-wave
approximation (RWA), we map the driven transport problem to a
static one which we solve analytically.
In doing so, a particular challenge represents the consistent RWA
treatment of the connecting leads in the presence of ac fields.

\section{The double-dot model}
\begin{figure}[b]
  \centering
  \includegraphics{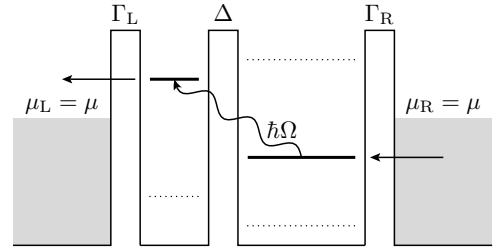}
  \caption{Level structure of the asymmetric double quantum dot in a pump
  configuration.  The solid lines mark the relevant levels $|1\rangle$ and
  $|2\rangle$ with the energies $\pm\epsilon_0/2$.
  The arrows indicate the dominating scattering process.}
  \label{fig:model}
\end{figure}%
We consider the setup sketched in Fig.~\ref{fig:model}
described by the time-dependent Hamiltonian $H(t) = H_\mathrm{dots}(t)
+ H_\mathrm{leads} + H_\mathrm{contacts}$, where the different
contributions correspond to the quantum dots, the leads, and the
tunneling coupling to the respective lead.
We disregard interaction and spin effects and assume that intra-dot excitations
do not play a role such that each dot is well described by a single energy
level.  Then, the double quantum dot Hamiltonian reads
\begin{equation}
  H_\mathrm{dots}(t)
  =
  - \frac{\Delta}{2} (c_1^\dagger c_2+c_2^\dagger c_1 )
  + \frac{\epsilon(t)}{2} (c_1^\dagger c_1-c_2^\dagger c_2 ) ,
\label{Hdots}
\end{equation}
where the fermion operators $c_{1,2}$ and $c_{1,2}^\dagger$ annihilate
and create an electron in the left and the right dot, respectively.
The on-site energy difference $\epsilon(t) = \epsilon_0 + A\cos(\Omega
t)$ is determined by the static internal bias $\epsilon_0$, the driving
amplitude $A$, and the frequency $\Omega$.  Typical driving frequencies
range up to 100\,GHz \cite{VanderWiel1999a} such that the
wavelength exceeds the size of the setup and, thus, the implicitly
assumed dipole approximation is well justified.

The leads are modeled as ideal electron gases, $H_\mathrm{leads} =
\sum_{q} \epsilon_q ( c_{\L q}^\dagger c_{\L q} + c_{\R q}^\dagger c_{\R q})$,
where $c_{\ell q}^\dagger$ creates an electron in lead $\ell = \L,\R$.
The tunneling Hamiltonian
\begin{equation}
H_\mathrm{contacts}
= \sum_{q} \left( V_{\L q} c^{\dagger}_{\L q} c_1
  + V_{\R q} c^{\dag}_{\R q} c_2 \right)
  + \mathrm{H.c.}
\end{equation}
establishes the contact between the dot levels and the respective
lead.
Below, we shall assume within a so-termed wide-band limit that the
coupling strengths $\Gamma_{\ell}=2\pi\sum_q |V_{\ell q}|^2
\delta(\epsilon-\epsilon_q)$, $\ell=\L,\R$, are energy independent.
To specify the dynamics, we choose as an initial condition for the
lead electrons a grand canonical ensemble at temperature $T$ and
chemical potentials $\mu_{\L,\R}$.
The influence of lead $\ell$ is fully determined by the lesser Green
functions $ g_{\ell q}^{<}(t,t') = (i/\hbar)\langle c_{\ell
  q}^\dagger(t') c_{\ell q}(t)\rangle$ and the tunnel matrix elements
$V_{\ell q}$ \cite{Meir1992a}.  More precisely, these quantities enter
the expressions for the current and the noise in form of the
correlation function
\begin{equation}
  \label{xi2}
  \langle \xi_\ell^\dagger(t-\tau) \xi_\ell(t) \rangle
  = \frac{\Gamma_\ell}{2\pi\hbar^2} \int d\epsilon\,
  e^{-i\epsilon \tau /\hbar} f_\ell(\epsilon)
\end{equation}
of the fermionic noise operator $\xi_\ell = -(i/\hbar)\sum_q V_{\ell
  q}^* c_{\ell q}$, where $f_\ell(\epsilon) =
(1+\exp[(\epsilon-\mu_\ell)/k_BT])^{-1}$ denotes the Fermi function
\cite{Kohler2005a}.  Henceforth, we consider the case of zero bias
voltage with
both chemical potentials located midway between the dot
levels $\pm\epsilon_0/2$, i.e., $\mu_\L=\mu_\R=0$.

\section{Resonant electron pumping}
For harmonic driving, the Hamiltonian $H(t)$ obeys time-reversal
symmetry and, hence, each individual scattering process has a
time-reversed partner which occurs with the same probability.  Thus,
it is tempting to conclude that the net current of both partners and,
consequently, the pump current vanishes.
This, however, is not the case because the driving enables energy
non-conserving scattering.  In particular, there exist processes like
the one sketched in Fig.~\ref{fig:model}:
With the leads initially at equilibrium, an electron from the right lead
with energy below the Fermi surface is scattered into a state in the
left lead with energy above the Fermi surface. This process
contributes to the current.  By contrast, the time-reversed
process does not transport an electron because the respective initial
state is not occupied.  The net effect is transport of electrons
from the lower level to the higher level, i.e., from right to left.

None the less, the pump might vanish due to the presence of an additional
symmetry, such as generalized parity $(x,t) \to (-x,t+\pi/\Omega)$
which relates two scattering processes with identical \textit{initial}
energies.  Their contributions to the current cancel each other
\cite{Kohler2005a}.
With equally strong coupling to  the leads, $\Gamma_\L = \Gamma_\R =
\Gamma$, generalized parity is satisfied for $H(t)$ at zero internal
bias  $\epsilon_0=0$.  For finite bias $\epsilon_0 \neq 0$, however,
this symmetry is broken and, consequently a finite pump current
emerges.
Moreover, this pump current exhibits resonance peaks including
higher-order resonances \cite{Stafford1996a}.

Within our analytical approach, we focus on strongly biased
situations, $\epsilon_0 \gg \Delta$, and driving frequencies
close to the internal resonances of the double dot, $n\hbar\Omega =
(\epsilon_0^2 + \Delta^2)^{1/2} \approx \epsilon_0$.  In this regime, the
dynamics of the dot electrons is dominated by the second term of the
Hamiltonian \eqref{Hdots} while the tunneling contribution, which is
proportional to $\Delta$, represents a perturbation.  Consequently, a
proper interaction picture is defined by the transformation
$ U(t) = \exp[-{\frac{i}{2}(c_1^\dagger c_1 - c_2^\dagger c_2 ) \phi(t)}]$
with the time-dependent phase
\begin{equation}
   \phi(t) = n\Omega t + \frac{A}{\hbar\Omega}\sin(\Omega t) .
\end{equation}
This yields the double-dot interaction-picture Hamiltonian $\widetilde
H_\mathrm{dots}(t) = U^\dagger(t) H_\mathrm{dots}(t) U(t) -i\hbar U^\dagger(t)
\dot U(t)$.  The transformation $U(t)$ has been constructed such that
$\widetilde H(t)$ obeys the time-periodicity of the original Hamiltonian
\eqref{Hdots} while all its other energy scales are significantly
smaller than $\hbar\Omega$.  Thus, we can separate time scales and replace
$\widetilde H_\mathrm{dots}(t)$ within a rotating-wave approximation (RWA)
by its time average
\begin{equation}
\label{Hdots.eff}
\bar H_\mathrm{dots}
= - \frac{\Delta_\eff}{2}( c_1^\dagger c_2  + c_2^\dagger c_1 )
  - \frac{\delta}{2}( c_1^\dagger c_1 - c_2^\dagger c_2)
\end{equation}
with $\delta = n\hbar\Omega - \epsilon_0$ and the effective tunnel
matrix element
\begin{equation}
  \label{Delta.eff}
  \Delta_\eff = (-1)^{n} J_n(A/\hbar\Omega)\Delta,
\end{equation}
where $J_n$ is the $n$th order Bessel function of the first kind.

While the lead Hamiltonian is unaffected by the transformation $U(t)$, the
tunneling Hamiltonian acquires a time-dependence, $\widetilde
H_\mathrm{contacts}(t) = \sum_{q} V_{\L q} c_{\L q}^\dagger c_1
e^{-i\phi(t)/2} + V_{\R q} c_{\R q}^\dagger c_2 e^{i\phi(t)/2} $.
The lead elimination along the lines of Ref.~\cite{Meir1992a}, but
here for a time-dependent contact Hamiltonian, reveals that the
influence of the leads is no longer determined by the noise operators
$\xi_{\ell}$ but rather by
$\eta_{\L/\R}(t) = e^{\pm i\phi(t)/2} \xi_{\L/\R}(t)$.
Its correlation function
$\langle\eta_{\L/\R}^\dagger(t-\tau) \eta_{\L/\R}(t)\rangle$
depends not only on the time-difference $\tau$, but also explicitly on $t$.
The latter time-dependence is $2\pi/\Omega$-periodic and is therefore
much faster than all other time scales.  Hence, we can replace
within the RWA the correlation function of $\eta_{\L,\R}$ by its
$t$-average
\begin{equation}
\label{eta2}
\overline{ \langle\eta_\ell^\dagger(t-\tau) \eta_\ell(t)\rangle }
= \frac{\Gamma}{2\pi\hbar^2} \int d\epsilon\,
  e^{-i\epsilon\tau/\hbar} f_{\ell,\eff}(\epsilon) ,
\end{equation}
where
\begin{equation}
  \label{feff}
  f_{\L/\R,\eff}(\epsilon)
  = \sum_{k=-\infty}^{\infty}
  J_k^2\Bigl(\frac{A}{2\hbar\Omega}\Bigr)
  f_{\L/\R}\Bigl(\epsilon+\Bigl[k \mp \frac{n}{2}\Bigr]\hbar\Omega\Bigr)
\end{equation}
can be interpreted as an effective electron occupation number of the
levels in lead $\ell$.  At zero temperature, it exhibits steps at
$\epsilon=\mu_{\ell}+(k\mp n/2)\hbar\Omega$ and is constant elsewhere.

The RWA provides a mapping of the originally time-dependent transport
problem to a static one with renormalized parameters.  This problem,
in turn, can be solved by standard procedures: Both the current and
the noise power can be expressed in terms of the transmission probability
$T(\epsilon)$ of an electron with energy $\epsilon$.  For a two-level
system in the wide-band limit, one obtains
\begin{equation}
\label{T}
T(\epsilon)
= \Gamma^2 |G_{12}(\epsilon)|^2
= \frac{\Gamma^2\Delta_\eff^2}{|(2\epsilon-i\Gamma)^2 -
       \Delta_\eff^2 - \delta^2|^2} .
\end{equation}
Then, the current defined as the change of the charge in the, e.g.,
left lead, is given by the Landauer-like formula
$I = (e/2\pi\hbar)\int d\epsilon\, T(\epsilon)[f_{\L,\eff}(\epsilon) -
f_{\R,\eff}(\epsilon)]$; the corresponding expression for
the noise power reads \cite{Blanter2000a}
\begin{equation}
\label{S}
\begin{split}
S =  \frac{e^2}{\pi\hbar} \int d\epsilon\, &T(\epsilon)\Big\{ \sum_{\ell=\L,\R}
f_{\ell,\eff}(\epsilon)[1-f_{\ell,\eff}(\epsilon)]
\\
+& [1-T(\epsilon)][f_{\L,\eff}(\epsilon)
                -f_{\R,\eff}(\epsilon)]^2 \Big\} .
\end{split}
\end{equation}
Note that in the presence of a driving field, even at zero temperature, the
electron occupation $f_{\ell,\mathrm{eff}}$ is not a simple step function
and, thus, also the term in the first line of Eq.~\eqref{S} contributes to
the noise power.
A convenient measure for the \textit{relative} noise strength is the Fano
factor $F=S/2eI$ which characterizes the noise with respect to the shot noise
level given by $S=2eI$ \cite{Blanter2000a}.

For the remaining evaluation of the energy integrals, it is important to
note that the transmission \eqref{T} is practically zero for $\epsilon^2
\gtrsim \Delta_\mathrm{eff}^2 + \Gamma^2 + \delta^2$.
Thus, for $\hbar\Omega \gtrsim \Delta_\mathrm{eff}, \Gamma, \delta$, the
effective electron occupation \eqref{feff} is constant in the relevant
energy range and can be replaced by its value at $\epsilon=0$.  One
obtains close to the $n$th resonance
\begin{align}
\label{Ipump}
I^{(n)}
= {} & \frac{e\Gamma}{2\hbar}\,
    \frac{\lambda_{n}\Delta_\eff^2}{\Delta_\eff^2 +
    \Gamma^{2} + \delta^{2}} ,
\\
\label{Spump}
S^{(n)}
= {} &\frac{e^2\Gamma}{2\hbar}\,
  \frac{\lambda_{n}^{2}\Delta_\eff^2[2(\Gamma^2+\delta^{2})^{2}
    -\Delta_\eff^2(\Gamma^2-3\delta^{2})
    +\Delta_\eff^4]}{(\Delta_\eff^2+\Gamma^2+\delta^{2})^3}
  \nonumber
  \\ &
  + \frac{1-\lambda_{n}^{2}}{\lambda_{n}} \, eI^{(n)},
\end{align}
where $\lambda_{n} = f_{\L,\eff}(0)-f_{\R,\eff}(0)
= \sum_{|k|\le n/2} J_k^2(A/2\hbar\Omega)$ with $|\lambda_n|\leq 1$.
Quite remarkably, for resonant driving ($\delta=0$), the pump current
assumes a \textit{maximum} while the noise power $S$ generally assumes
a local \textit{minimum}; cf.\ Fig.~\ref{fig:omega}(a).  This results
in an even more pronounced minimum for the Fano factor.

\section{Floquet transport theory}
Before developing an optimization strategy, we
corroborate our analytical results by an exact numerical calculation within
Floquet transport theory \cite{Kohler2005a}:
Starting from the Heisenberg equations of motion for the
annihilation operators for both the lead and the dot electrons, one
eliminates the lead operators and thereby obtains for the electrons on the
dots a reduced set of equations.  These are solved with the help of
the retarded Green function obeying $[i\hbar d/dt - \mathcal{H}(t) +
i\Gamma/2 ] G(t,t') = \delta(t-t'),$ where $\mathcal{H}(t)$ is the
single-particle Hamiltonian corresponding to the double-dot Hamiltonian
\eqref{Hdots}.
The coefficients of the equation of motion for $G(t,t')$ are
$2\pi/\Omega$-periodic and, consequently, its solution can be
constructed with the help of the Floquet ansatz $|\psi_\alpha(t)\rangle =
\exp[(-i\epsilon_\alpha/\hbar-\gamma_\alpha)t]|\phi_\alpha(t)\rangle$.  The
Floquet states $|\phi_\alpha(t)\rangle$ obey
the eigenvalue equation
$[\mathcal{H}(t) - i\Gamma/2 - i\hbar d/dt ] |\phi_\alpha(t)\rangle
= (\epsilon_\alpha - i\hbar\gamma_\alpha)|\phi_\alpha(t)\rangle$.
Its solution allows to construct the retarded Green function $G(t,t')
= -(i/\hbar) \sum_\alpha
|\psi_\alpha(t)\rangle\langle\psi_\alpha^+(t')| \Theta(t-t')$.
Finally, one obtains for the pump current a convenient Landauer-like
expression with an additional sum over the sidebands
\cite{Camalet2003a, Kohler2005a}.
Since the symmetrized noise correlation function $S(t,t') =
\langle [I(t),I(t')]_+\rangle$ depends explicitly on both times, we
characterize the noise by the time-average of its zero-frequency component,
$S = (2\pi/\Omega)\int_0^{2\pi/\Omega} dt \int d\tau S(t,t-\tau)$.

\begin{figure}[t]
\includegraphics{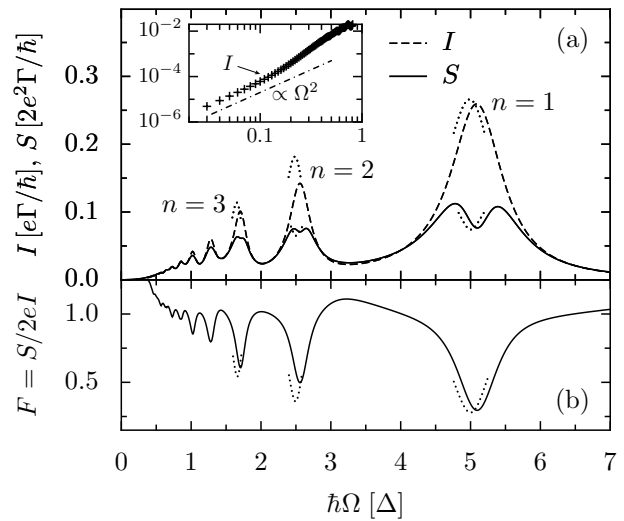}
\caption{\label{fig:omega}
(a) Pump current $I$ (dashed line) and its noise power $S$ (solid) at $k_BT=0$
as a function of the driving frequency for coupling strength
$\Gamma=0.3\Delta$, driving amplitude $A=3.7\Delta$, and internal bias
$\epsilon_0=5\Delta$.  The dotted lines mark the analytical results
\eqref{Ipump} and \eqref{Spump}.
Inset: Blow-up of the lower-left corner demonstrating that $I\propto\Omega^2$.
(b) Corresponding Fano factor.}
\end{figure}%
Figure \ref{fig:omega}(a) depicts the numerically evaluated pump current and
its noise power.  For proper cooling, thermal excitations do not play a
significant role.  Therefore, we consider zero
temperature only.  We find that the current exhibits peaks located at the
resonance frequencies $\Omega = (\epsilon_0^2 +\Delta^2)^{1/2}/n\hbar$.
This agrees well with our analytical results (dotted lines), albeit
the RWA predicts the location of the current maxima only to
zeroth order in $\Delta$, i.e., at the slightly shifted frequencies
$\Omega=\epsilon_0/n\hbar$.
In the adiabatic limit, the pump current vanishes proportional to $\Omega^2$.
For the chosen parameters, the noise power $S$ possesses
clear minima, each accompanied by two maxima.  In the vicinity
of the resonance, the noise is considerably below the shot noise level
$2eI$; cf.\ Fig.~\ref{fig:omega}(b).
This feature is notably pronounced at the first resonance.  Far from
the resonances, the current becomes smaller and the Fano factor is
close to $F=1$.
The comparison of the numerically exact results with the current \eqref{Ipump} and the
noise power \eqref{Spump} [dotted lines in Fig.~\ref{fig:omega}(a)] leads to
the conclusion, that the RWA predicts both the current maxima and the noise
minima sufficiently well to employ these expressions for a parameter
optimization towards low-noise pumping.

\section{Tuning the electron pump}
We have already seen that the condition of large current and low noise is
met at the internal resonances of the biased double-dot setup.  Thus, we
can restrict the search for optimal parameters to resonant driving.
As a figure of merit for the noise strength we employ the
Fano factor for $\delta=0$
\begin{equation}
\label{Fpump}
  F^{(n)}
  = \frac{S^{(n)}}{2e I^{(n)}}
   = \frac{1}{2\lambda_{n}} - \frac{\lambda_n}{2}
     \frac{\Gamma^2(3\Delta_\eff^2-\Gamma^2)}{(\Delta_\eff^2+\Gamma^2)^2},
\end{equation}
which is a function of $\lambda_n$ and $\Delta_\mathrm{eff}/\Gamma$.
The second term is minimal for $\Delta_\mathrm{eff}/\Gamma =
\sqrt{5/3}$, yielding $F^{(n)} = 1/(2\lambda_n) - 9\lambda_n/32$.
Thus, the optimal Fano factor is assumed for $\lambda_n=1$ and reads
$F_\mathrm{opt} = 7/32 \approx 0.219$.
\begin{figure}[t]
\includegraphics{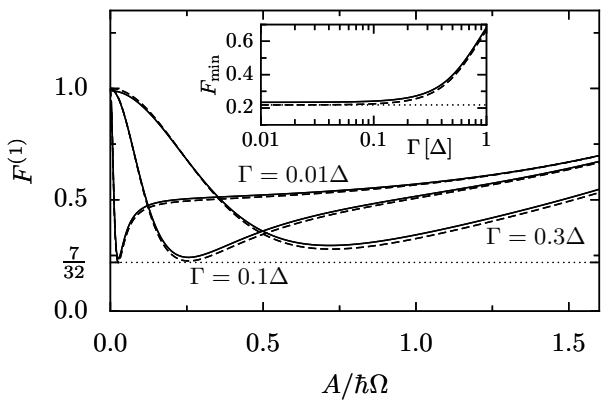}
\caption{\label{fig:optimizing}
Fano factor $F$ at the first resonance for various coupling strengths
$\Gamma$.  The exact Floquet calculation (solid lines) is compared with the
RWA for $\delta=0$ (dashed).  The inset depicts the minimal Fano
factor in dependence of $\Gamma$ for $\epsilon_{0}=5\Delta$ at the
resonance $\Omega=\sqrt{\Delta^2+\epsilon_0^2}/\hbar$.  The dotted lines
mark the optimal Fano factor $F_\mathrm{opt}=7/32$.}
\end{figure}%

In the following, we restrict ourselves to the prime resonance
($n=1$) for which $\Delta_\mathrm{eff} =
J_1(A/\hbar\Omega)\Delta$ and $\lambda_1 = J_0^2(A/2\hbar\Omega)$
\cite{on_generalization}.  Then, the value $\lambda_1=1$ is assumed
for $A=0$ which means
$\Delta_\mathrm{eff}=0$; this unfortunately implies a
vanishing current \eqref{Ipump}.  Therefore, the central question is
whether it is possible to find a driving amplitude providing on
the one hand an appreciably large pump current, while on the other
hand yielding a noise level close to $F_\mathrm{opt}$.  The
numerical results depicted in Fig.~\ref{fig:optimizing} indeed
suggest this possibility: The Fano factor is close to the optimal value
already for a finite amplitude.
A closer investigation reveals that the location
of the minimum corresponds to $\Delta_\mathrm{eff} =
J_1(A/\hbar\Omega)\Delta = \sqrt{5/3}\,\Gamma$, in compliance with our
analytical considerations.  In particular,
the minimum is shifted towards smaller values of
$A/\hbar\Omega$ for weaker coupling $\Gamma$.  Moreover, the RWA
solutions \eqref{Ipump} and \eqref{Spump} agree very well with the
numerically exact results,
although they slightly underestimate the noise.  This
discrepancy diminishes as $\Omega^{-2}$ (not shown).

The data also reveal that in the interesting regime, the ratio
$A/\hbar\Omega$ is considerably smaller than 1 and, hence, we can employ the
approximations $J_0(x)\approx 1-x^2/4$ and J$_1(x)\approx x/2$ valid for small
arguments.  It is now straightforward to obtain to lowest order in
$A/\hbar\Omega$ the expressions $\Delta_\mathrm{eff} =
A\Delta/2\hbar\Omega$ and $F^{(1)} = 7/32 + (5A/16\hbar\Omega)^2$.
For instance, choosing $A=0.3\hbar\Omega$, the noise level lies merely
5\% above $F_\mathrm{opt}$ and the condition $\Delta_\mathrm{eff} =
\sqrt{5/3}\,\Gamma$ corresponds to $\Gamma\approx 0.1\Delta$, i.e., to
weak dot-lead coupling.  This estimate is confirmed by the inset in
Fig.~\ref{fig:optimizing} which, in addition,
demonstrates that $F \approx F_\mathrm{opt}$ for $\Gamma\lesssim
0.1\Delta$.   For such a small coupling $\Gamma$, interaction-induced
electron-electron correlations typically play a minor role.

In the experiment of Ref.~\cite{VanderWiel1999a}, a typical inter-dot
coupling is $\Delta = 50\mu\mathrm{eV}$.  Then, an internal bias
$\epsilon_0=5\Delta$ corresponds to the resonance frequency
$\Omega=5\Delta/\hbar \approx 2\pi\times 60\mathrm{GHz}$.  Tuning the
lead coupling to $\Gamma=0.1\Delta$ results in an optimized pump current of the
order $200\mathrm{pA}$ with a Fano factor $F\approx0.23$.

\section{Conclusions}
The analysis of ac-driven, asymmetric double quantum dots demonstrates
that optimal pumping is achieved beyond the adiabatic regime.  In
particular, the ideal \textit{modus operandi} requires a large
internal bias at resonant driving in combination with a strong
inter-dot coupling $\Delta\gtrsim 10\Gamma$.  The resulting pump
current then assumes a maximum while, interestingly enough, the
(absolute) noise power assumes at the
same time a minimum such that the Fano factor becomes remarkably small.
In order to systematically tune the pump into a low-noise regime, we
have derived analytical
expressions for both the current and its noise power.  The comparison
with the numerically exact solution fully confirms the validity of this
approach.  Our findings convincingly suggest that coupled quantum dots
are ideal for pumping electrons effectively and reliably at a low noise level.

We acknowledge discussions with S. Camalet, G.-L. Ingold, and
J. Lehmann.
This work has been supported by the DFG
through Graduiertenkolleg 283 and SFB 486.


\begin{thebibliography}{10}
\bibitem{on_generalization}
At higher resonances, the same optimal Fano factor can be reached with a larger
  driving amplitude.
\end{thebibliography}
\end{document}